\documentclass[5p,onecolumn,10pt]{elsarticle}
\usepackage[title]{appendix}
\usepackage{amsmath}
\usepackage{hyperref}
\usepackage{soul}
\usepackage{lineno}

\usepackage{multirow}
\usepackage{makecell}
\usepackage{epstopdf}
\usepackage{amsmath}%
\usepackage{bm}
\usepackage{amsfonts}%
\usepackage{amssymb}%
\usepackage{graphicx}
\usepackage{mathrsfs}
\usepackage{algorithm}
\usepackage[noend]{algpseudocode}
\usepackage{tikz-cd}
\usepackage{url}
\usepackage{arydshln}
\usepackage{fancyhdr}
\usepackage{array}
\usepackage{lineno}

\modulolinenumbers[5]

\makeatletter
\def\BState{\State\hskip-\ALG@thistlm}
\makeatother

\usepackage{comment} 
\usepackage{subfig}
\usepackage{hyperref}
\hypersetup{colorlinks=true,allcolors=blue}
\usepackage{hypcap}

\newcolumntype{C}{>{\centering\arraybackslash} m	 } 

\newcommand{\svnidlong}[4]{}%
\newcommand{\ch}[1]{{\color{black}{#1}}}
\newcommand{\x}[1]{{\color{blue}{#1}}}

\begin{document}
\baselineskip 11pt

\begin{frontmatter}

\title{A Framework for Simulating the Path-level Residual Stress in the Laser Powder Bed Fusion Process}

\author[1]{Xin Liu}
\ead{xin@intact-solutions.com}

\author[1]{Xingchen Liu\corref{cor1}}
\ead{xliu@intact-solutions.com}

\author[2]{Paul Witherell}
\ead{paul.witherell@nist.gov}

\cortext[cor1]{Corresponding author}
\address[1]{Intact Solutions, Inc.}
\address[2]{National Institute of Standards and Technology}

\begin{abstract}

Laser Powder Bed Fusion (LPBF) additive manufacturing has revolutionized industries with its capability to create intricate and customized components. The LPBF process uses moving heat sources to melt and solidify metal powders. The fast melting and cooling leads to residual stress, which critically affects the part quality. Currently, the computational intensity of accurately simulating the residual stress on the path scale remains a significant challenge, limiting our understanding of the LPBF processes. 
 
This paper presents a framework for simulating the LPBF process residual stress based on the path-level thermal history. Compared with the existing approaches, the path-level simulation requires discretization only to capture the scanning path rather than the details of the melt pools, thus requiring less dense mesh and is more computationally efficient. We develop this framework by introducing a new concept termed effective thermal strain to capture the anisotropic thermal strain near and around the melt pool. We validate our approach with the high-fidelity results from the literature. We use the proposed approach to simulate various single-island scanning patterns and layers with multiple full and trimmed islands. We further investigate the influence of the path-level thermal history and the layer shape on the residual stress by analyzing their simulation results.

\end{abstract}


\end{frontmatter}

\section{Introduction}

Laser Powder Bed Fusion (LPBF) is a form of metal additive manufacturing (AM) that employs lasers as moving heat sources to melt and solidify thin layers of metal powder along a predefined tool path, building the part layer by layer \cite{vock2019powders, bhavar2017review, singh2021powder}. LPBF has gained significant attention due to its ability to create lightweight, intricate structures. 
Despite its potential, the full exploitation of LPBF is hindered by the challenge of consistently predicting and controlling the quality of manufactured components and material performance \cite{liu2016homogenization}. Complex thermal-mechanical processes give rise to uncertainties such as undesirable deformations, material property variations, and residual stress, which pose significant problems, especially in industries where reliability and precision are paramount. The inability to accurately predict and mitigate these issues has restricted the application of AM in critical areas. The complexity is compounded by scanning paths, as different paths lead to distinct thermal histories and residual stresses, emphasizing the need to understand their influence on material properties.

The LPBF process involves many complex multi-physics and multi-phase phenomena, including the melting of metal powders and remelting of solidified metal, the fluid dynamics and heat transfer of the molten metal, and the microscopic grain structure resulting from the solidification of liquid metals, and so on. Capturing all these effects from the first principle can be extremely computationally expensive, even if only for a small domain. For example, it takes up to 4000 minutes to simulate a single scan melt pool on a $390 \times 210 \times 50 \mu m^3$ domain~\cite{luthi2023adaptive}. To simulate the LPBF residual stress, researchers made various assumptions to simplify the model. For example, the liquid or gas phase is replaced by an equivalent solid volume or ignored~\cite{moges2019review}. The simulation is simplified and usually implemented by the finite element method with fine discretization. The discretization need to be fine enough to capture the melt pool shape. For simplicity, we will refer this approach as the "voxel-based approach" in the present paper. Despite the assumptions and simplifications, the voxel-based approach is still expensive and so far is only applicable to a small domain. For example, the simulation of a single layer whose size is 2 mm by 2 mm in the literature takes more than 1 day~\cite{patil2021benchmark}. 

To efficiently simulate the residual stress, researchers proposed agglomeration approaches such as the inherent strain method~\cite{liang2018modified,liang2021incorporating} that completely bypass the laser scanning path. This method is a two-step approach. Firstly a full-scale simulation is conducted on a small sample domain to extract the effective ``inherent strain''. Then, multiple powder layers are agglomerated as a single ``superlayer'' where the simulation is conducted as a superlayer-wise activation process. The inherent strain is applied on each superlayer sequentially. This agglomeration approach is very efficient, however, it can no longer predict the effect of the scanning paths. 
\ch{The correlation between the thermal history and the scanning path especially the sequence of melting and solidification is ignored.}
This is particularly problematic for the boundary region of a layer, as well as layers with intricate details, where the influence of the scanning path is more pronounced.

In our previous work, we developed PBF-CAPL to capture the thermal histories of the LPBF process~\cite{liu2024scalable} on the path level. This approach efficiently captures the influence of the scanning paths on the thermal history. PBF-CAPL differs from the conventional finite element approaches as PBF-CAPL only requires discretization on the path level. \ch{The discretized elements of PBF-CAPL are some line segments along the scanning path.} A lumped model is defined on such line segments to model the convection-conduction-radiation problem. We validated this approach by comparing the experimental and simulation melt pool length. Readers might refer to the appendix and our previous paper~\cite{liu2024scalable} for more details about PBF-CAPL.  


Multiple studies~\cite{parry2016understanding, denlinger2017thermomechanical, chen2019effect, zhang2020scanning} suggested that the residual stress in the LPBF layer is anisotropic: the stress is more dominant in the laser scanning direction than the transverse direction in the powder layer plane. We will use the term ``anisotropic stress'' of the LPBF process for simplicity. Such anisotropic stress is due to the thermal gradient (spatial temperature gradient) around the melt pool being highly anisotropic~\cite{parry2016understanding}. Because the PBF-CAPL uses the lumped element along the scanning path, it does not capture such anisotropic thermal gradients on the path scale. Consequently, PBF-CAPL results can not be applied directly to the conventional finite element-based mechanical simulations.

\begin{figure}[H]
    \centering
    \includegraphics[width = 0.8\textwidth]{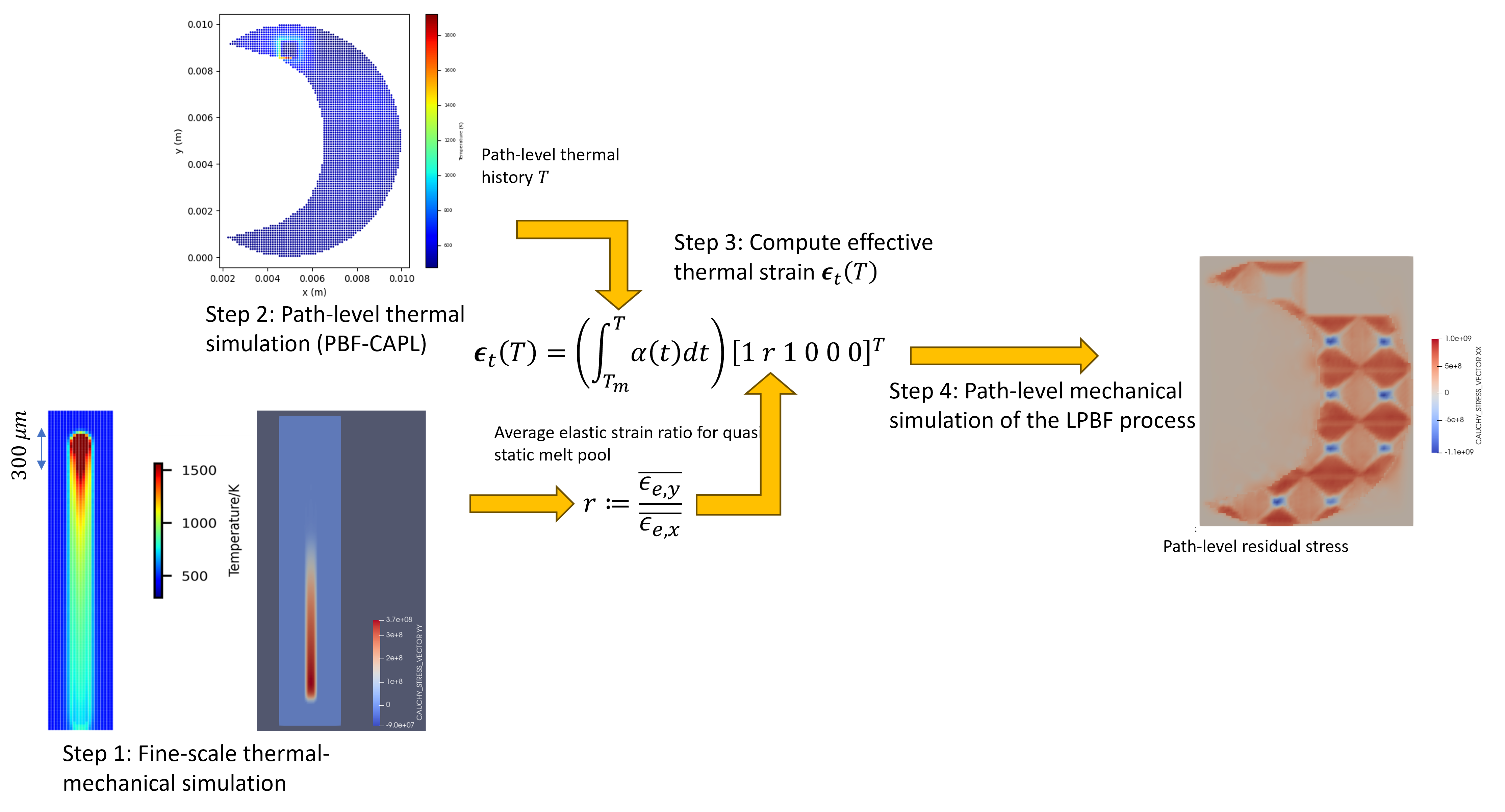}
    \caption{Outline of the path-level LPBF simulation}
    \label{fig:outline}
\end{figure}

In the present paper, we developed a novel thermomechanical simulation approach for LPBF based on the path-level thermal history. The new approach requires a coarser discretization compared to those used in conventional voxel-based approaches. An overview of the new approach is shown in Figure \ref{fig:outline}. \ch{ Firstly we compute the residual stress of around the steady-state melt pool from a fine-scale simulation to capture the anisotropic stresses. Secondly, we use a path-level thermal simulation to obtain the thermal history for the given scanning path. Then we utilize the fine-scale results and the path-level thermal history to compute an ``effective thermal strain'' which reflects the anisotropic stresses. } The effective thermal strain is an anisotropic strain that is to be used in the path-level mechanical simulation. Lastly, the effective thermal strain is applied to the path-level mechanical simulation to simulate the LPBF path-level residual stress.

Our main contribution is to provide an approach to simulate the residual stress on the path level. We develop this new approach with an effective thermal strain used to address the anisotropic stresses in the scanning and transverse directions. The new approach is more efficient compared with the conventional voxel-based approach which requires the discretization to capture the details of the melt pool. The discretization used in the proposed approach only needs to capture the scanning paths. Thanks to the better efficiency, we can simulate the LPBF scanning process on larger and more complex layers. We further investigate the effects of layer shapes and scanning paths on the residual stress. To the best of the authors' knowledge, little work has been done to investigate the influence of the laser scanning path especially the sequence of the melting and solidification on the residual stress using simulations. 

The rest of the paper is organized as follows. We review the related work in Section 2. In Section 3, we formulate and validate a new path-level mechanical simulation framework for the LPBF process. In Section 4, we present the simulation results of different island checkerboard patterns, including single island tests and single layer tests which include multiple full or trimmed islands. We discuss the effect of the scanning path on the path-level residual stress based on our simulation results. In the last section, we conclude our work in the present work and discuss the possible future work.
\section{Related Work}
\subsection{High-fidelity meltpool dynamics simulation}

Simulating the LPBF process is a complex multiphysics and multiphase problem if all different physics and interactions between phases are modeled. For example, simulating melt pool dynamics might involve solidified metal powders~\cite{gourdin1986dynamic}, fluid dynamics of molten metal flow~\cite{queva2020numerical}, particle dynamics of metal powders~\cite{bidare2018fluid}, and gas dynamics of evaporated metal gas \cite{matthews2017denudation}, all of which interact with each other~\cite{michopoulos2018multiphysics}. Such simulation has been conducted and it provided valuable insights into the LPBF process, however, it is impossible to use it for practical applications due to the extremely high computational cost.

\subsection{Voxel-based simulation}
The voxel-based simulation is commonly used for simulations of the LPBF process beyond melt pools. The voxel-based simulation is based on multiple assumptions and techniques which are used to simplify the modeling. Firstly, the fluid flow upon melting is not explicitly modeled~\cite{ganeriwala2021towards}. All materials, including powder, liquid, and solid materials, are represented as a continuum with modified material. The material is modeled by some material properties such as density, Young's modulus, and Poisson ratio. These material properties are modeled as functions of temperature. Under this assumption, modeling of grain structure, powder spatial configuration, and liquid melt pool are no longer needed. Many papers have been published on simulating the melt pool shape using these assumptions, often with fine spatial and temporal discretization~\cite{parry2016understanding,luo2018survey,cao2021novel}.

Weak coupling between thermal and mechanical simulation is a common assumption used in voxel-based approach. This assumption is based on the observation that the deformation during the LPBF process is small. Specifically, it suggests the mechanical analysis relies on the thermal analysis but the thermal analysis is independent of the mechanical analysis~\cite{luo2018survey}.
Therefore, we only need to consider the temperature changes that drive thermal expansion/shrinkage in the mechanical analysis. This assumption also allows the asynchronous thermal analysis before the mechanical analysis.

\begin{figure}
    \centering
    \includegraphics[width=0.4\textwidth]{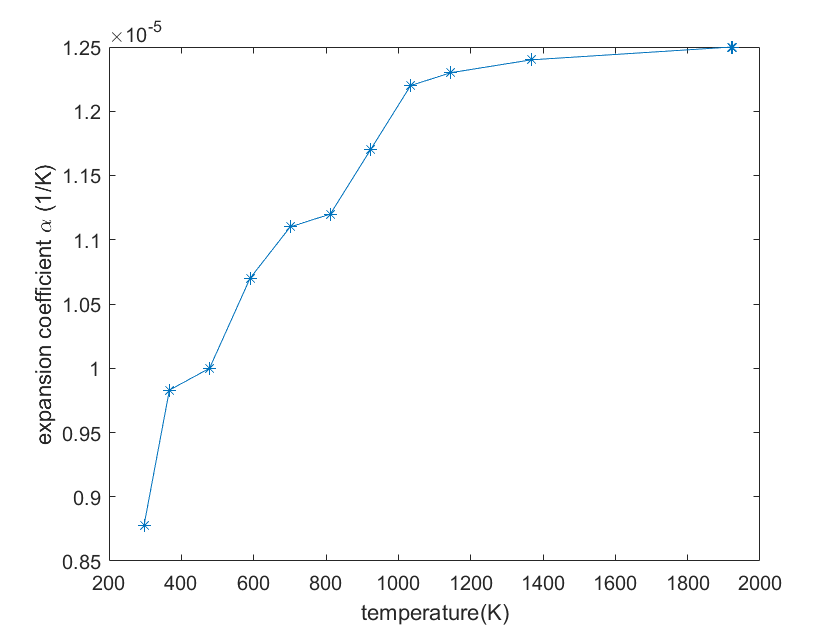}
    \caption{The expansion coefficient of Ti6Al4V as a function of temperature.}
    \label{fig:expansion_coefficient}
\end{figure}

In the mechanical analysis, the total strain $\epsilon_{total}$ in the LPBF process is decomposed as:
\begin{align}\label{eq:strains}
    \bm{\epsilon_{total}} = \bm{\epsilon_{t}} + \bm{\epsilon_{e}} + \bm{\epsilon_{p}}
\end{align}
where $\bm{\epsilon_{t}} + \bm{\epsilon_{e}} + \bm{\epsilon_{p}}$ are the thermal, elastic and plastic strain, respectively. Plastic $\bm{\epsilon_{p}}$ is obtained with the given plastic model, for example Von-Mises yield criterion. Thermal strain is obtained with a given thermal history in the form $\bm{\epsilon_t}(T) = \epsilon_t(T)[1, 1, 1, 0, 0, 0]^T$, where
\begin{align}
    \epsilon_t(T) = \int_{T_m}^T\alpha(t)dt 
\end{align}
The thermal expansion coefficient $\alpha(t)$ is a function of temperature $t$. Figure \ref{fig:expansion_coefficient} shows the $\alpha(t)$ for Ti6Al4V which we used in the present paper. The reference temperature $T_m$ is the metal melting point. All strain and stress states will be reset to zero when the temperature is above the melting point. The vector $[1,1,1,0,0,0]^T$ indicates the thermal strain is isotropic in the normal directions $x,y,z$ (the first three ones) and zero shear strain (the last three zeros). Despite employing isotropic thermal strain, the voxel-based approach has the capability to capture anisotropic stress with sufficient fine discretization. With adequate resolution, this approach can effectively capture the sequence of melting and solidification of the elements. The stress experienced by an element can vary depending on whether it is surrounded by already solidified elements or not. This variation can be reflected by the thermal gradient near the melt pool.

A technique called element birth and death is frequently used in simulating the LPBF process~\cite{yang2019residual}. This technique simulates the powder deposition process in the LPBF process. The powder deposition process is a layer-by-layer process. In each layer, the laser moves according to the scanning path to melt and solidify the metal powders. The direct modeling of this process is computationally expensive. Assuming the deformation during the scanning process is small, the element birth and death technique avoids such direct modeling by initializing and meshing the entire domain and updating the stiffness matrix of the mesh to mimic the deposition process. All elements are initially assigned a material property that is considered approximately void. This property will be switched to the normal value when powders are deposited in the layer to simulate the powder deposition process. For example, the element's Young's modulus may be initially assigned a very small value, such as $10^{-4}$ of the normal value ~\cite{denlinger2015residual, denlinger2017thermomechanical}. It will be switched to the powder's effective bulk material properties when powder deposition occurs on the element. 

Based on the above assumptions, the stepwise quasi-static mechanical simulation uses the equation below:
\begin{align}
    \bm{\nabla \sigma} &= 0 \\
    \bm{\sigma} &= \bm{C\epsilon_e} \\ 
    \frac{\partial \bm{u}}{\partial \bm{r}} &= \bm{\epsilon_{total}} = \bm{\epsilon_{t}} + \bm{\epsilon_{e}} + \bm{\epsilon_{p}} \\
    \bm{\epsilon_t}(T) &= \int_{T_m}^T\alpha(t)dt [1, 1, 1, 0, 0, 0]^T
\end{align}
The plastic strain $\bm{\epsilon_{p}}$ is computed with the plasticity model. In the present paper, we use the Von Mises yield surface. $\bm{u}, \bm{r}$ are the displacement and the coordinates respectively. The thermal strain is an integration of the thermal history $T$ according to the equation above. The thermal history $T$ comes from the thermal simulation which is a transient analysis driven by the moving heat source. The mechanical analysis is a step-wise quasi-static analysis. In each step, the mechanical analysis can take the corresponding thermal history from the thermal analysis. The laser energy input is typically modeled as a Gaussian~\cite{sow2020influence} or ring~\cite{grunewald2021influence} distribution, with the center moving along the predetermined scanning path. Many studies used the finite element method to implement this formulation~\cite{yan2015multiscale, dunbar2016experimental, cao2021novel}. The effect of scanning strategies in the LPBF process is investigated in papers~\cite{liu2022understanding, nadammal2021critical, chen2021prediction}. 

Such voxel-based thermomechanical simulation is computationally expensive and is limited to relatively short and simple scanning paths~\cite{luo2018survey}. This is because the discretization must be fine enough to capture the details of the melt pool. Since the melt pool size is small (melt pool width is usually around 100 $\mu$m for a typical LPBF process), the element from the discretization needs to be smaller (usually in the order of $10\mu$m). A small time step (usually in the order of $\mu$s) is needed because the time step needs to be compatible with the element size to ensure numerical stability. The small time step and element size make the computational cost of thermal-mechanical simulation prohibitively high for large-part simulation. For example, a 2 mm $\times$ 2 mm 3-layer structure may require up to 10 hours to simulate~\cite{cao2021novel}. The inefficiency is mainly due to the element size being small around the melt pool and the time step being small. The voxel-based approach scales with the fourth order of the element size: it necessitates three orders of refinement spatially for 3D space and one order temporally.


\subsection{Agglomeration approaches}
One of the main reasons for the high computational cost of conventional thermal-mechanical simulation is the need for dense spatial discretization and small time steps to model the LPBF process accurately. Agglomeration methods have been proposed to address this issue. Agglomeration methods use an intermediate, agglomerated model much larger than the scale of the spatial discretization elements. This approach can significantly reduce the computational cost of the simulation. ``Superlayer''~\cite{peng2018fast}, an artificial layer that composes multiple adjacent real powder layers, is commonly used as the agglomerated model. Layer-wise approaches, like the flash heating method~\cite{bayat2020part} or inherent strain method~\cite{liang2018modified,liang2021incorporating}, are proposed for better efficiency. In these methods, the scanning path is no longer considered. Instead, the superlayers are activated in order. Because of their fewer steps and less dense mesh, these methods are significantly more efficient compared to conventional thermal-mechanical simulations. However, by disregarding the scanning path, these approaches also lack the capability to predict the effects of scanning paths.

\ch{In the flash heating method, the moving heat source is replaced by an equivalent thermal load, which is activated superlayer-by-superlayer. Each superlayer is sequentially activated.} An equivalent heat is applied to the newly activated superlayer to conduct the thermal simulation. Then, the mechanical simulation is conducted by applying thermal strain, which is based on the temperature obtained from the thermal simulation to the activated superlayer.  

The inherent strain method was originally proposed for welding problems that are based on the assumption that residual stresses will completely relax after the welding process~\cite{ueda1979new}. However, this assumption is no longer valid for additive manufacturing, particularly in laser powder bed fusion, where non-uniform residual stress is trapped due to the complex scanning paths. Researchers have attempted to improve the inherent strain method to better account for the LPBF process. For example, a modified inherent strain method~\cite{liang2018modified} is proposed to extract the inherent strain vector from a fine-scale model. The work has been further improved to use representative volume elements to address periodically layer-wise rotating scanning paths~\cite{liang2021incorporating}. However, accurately predicting residual stress under more general scanning paths, complex geometry, and process parameters such as laser power and speed remains unaddressed.

\subsection{Path-level thermal simulation}

In our previous work, we developed a scalable PBF thermal simulation approach that we will refer to as Powder Bed Fusion Contact Aware Path Level (PBF-CAPL)~\cite{liu2024scalable}. This approach discretizes the simulation domain on the path level. We validate this approach against melt pool images captured with the co-axial melt pool monitoring system on the Manufacturing Metrology Testbed (AMMT) developed at the National Institute of Standards and Technology (NIST). Excellent agreements in the length and width of melt pools are found between simulations and experiments conducted on a custom-controlled LPBF testbed on a nickel-alloy solid surface.

\begin{figure}
    \centering
    \includegraphics[width=0.2\textwidth]{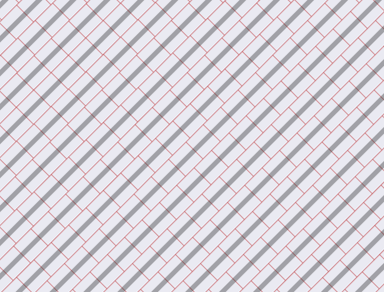}
    \caption{PBF-CAPL discretizes the scanning path (in black) into elements (in red). The element width is equal to the hatch distance.}
    \label{fig:capl-element}
\end{figure}

The discretization in PBF-CAPL is associated with the scanning path, where each element from the discretization corresponds to a segment along the scanning paths. see Figure \ref{fig:capl-element}. On each element, a lumped model is applied to simulate the convection-conduction-radiation problem driven by the moving heat source. Such path-level discretization is advantageous compared to the conventional approaches because a larger element size is used. As discussed previously, the simulation time is proportional to the fourth order of the inverse of the element size, using an element size of $100\mu$m is up to $5^4=625$ times faster compared to the conventional simulation which uses a $20\mu$m element. Interested readers might find more details about PBF-CAPL in the appendix.

\ch{However, the PBF-CAPL thermal history cannot be directly used in the voxel-based approaches. This is because the voxel-based approaches requires an accurate thermal gradient, but the thermal gradient is not directly available in the PBF-CAPL thermal history due to the lack of discretiztion and resolution in the direction transverse to the scanning path. The PBF-CAPL element width is too large to capture the thermal gradient around the melt pool: using a naive finite difference to compute the gradient is likely to give inaccurate results due to insufficient resolution along the transverse direction. } \ch{The importance of an accurate thermal gradient, however, is underscored by research findings that stress exhibits anisotropy in both the scanning and transverse directions along the scanning path~\cite{parry2016understanding,denlinger2017thermomechanical,chen2019effect,zhang2020scanning}. It has been pointed out that these anisotropic stresses stem from the anisotropic thermal gradient~\cite{parry2016understanding}.} In this paper, our approach aims to bridge the gap by leveraging the PBF-CAPL thermal history for voxel-based methods.

\section{Methodology of the path level mechanical simulation of LPBF}
\label{sec:methodolgy}

As discussed previously, voxel-based approach requires fine discretization to capture the thermal gradient around the melt pool. Consequently, the thermal history from PBF-CAPL cannot be directly applied because it lacks the resolution in the transverse direction. In our method, we propose a novel effective thermal strain to capture the anisotropic stresses. The proposed method includes four major steps: (1) Obtaining the desired anisotropic stress by simulating the residual stress of around a melt pool using the conventional voxel-based approach. (2) Conducting path-level thermal simulation by PBF-CAPL. (3) Constructing the effective thermal strain with the thermal history from the PBF-CAPL and the anisotropic stress from the voxel-based approach. (4) Conducting the path-level mechanical simulation by applying the effective thermal strain. \ch{We will show the details below with an example compared to the results by the voxel-based approach~\cite{parry2016understanding}. }

\ch{In the first step, we conduct the thermomechanical simulation of a steady-state melt pool to obtain the residual stress distribution with the voxel-based approach.} This step serves the purpose to obtain the ratio of the anisotropic stresses in the scanning and the transverse direction, which will be used to compute the effective thermal strain in the following steps. \ch{We choose the scanning path length to be 3 mm to ensure a quasi-steady state is reached. } The simulation use the weakly coupled voxel-based simulation. The thermal simulation is implemented with a homemade finite element code. The simulation and process parameters are given in table \ref{table:meltpool_process_parameters}. The mechanical simulation is implemented with the KratoMultiphysics~\cite{dadvand2010object,dadvand2013migration,ferrandiz2022kratos}. \ch{The melt pool temperature distribution and the stress are shown in Figure \ref{fig:fine_scale_result}.}

\begin{table}
\caption{Process parameters and material properties of Ti6Al4V~\cite{parry2016understanding}}\label{table:meltpool_process_parameters} 
\centering

\footnotesize
\begin{tabular}{ll}
 \hline
 Melting temperature &   1923 K \\
 Environmental temperature &   473 K \\
 Layer thickness & 40 $\mu$m \\
 Laser power & 82.5 W \\
 Laser speed & 0.5 m/s \\
 Environmental convection coefficient & 10 W/(m$^2$ K)\\
 Element size & 20 $\mu$m\\
 Laser spot size & 50 $\mu$m
\end{tabular}
\end{table} 

\begin{figure}
    \centering
    \includegraphics[width=0.3\textwidth]{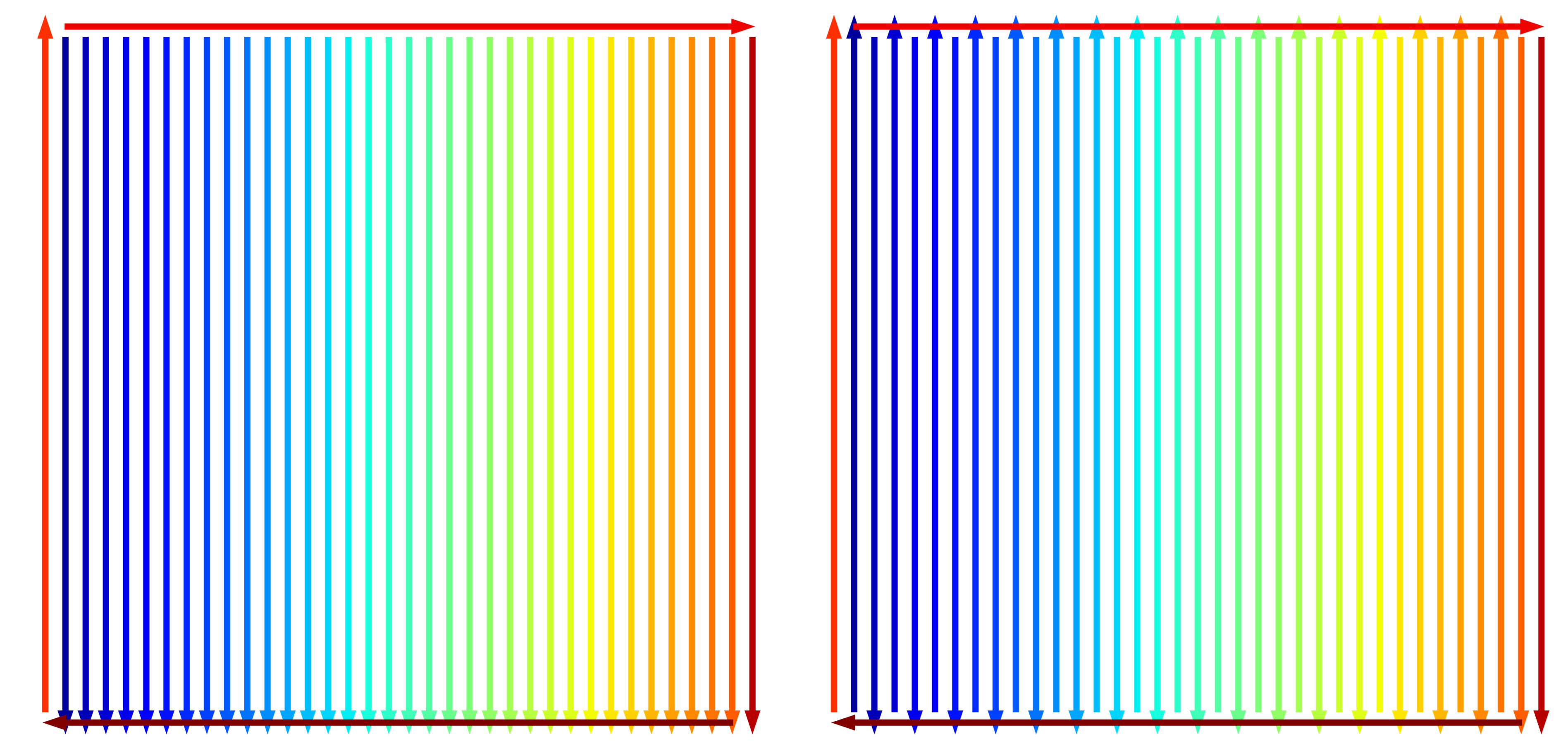}
    \caption{Scanning paths (unidirectional and alternating with a contour scanning) used by Parry et al.~\cite{parry2016understanding}. The laser start from blue to red. As shown by the color, the post contour scanning happens after the parallel scanning. }
    \label{fig:validation_path}
\end{figure}

\begin{figure}[H]
    \centering
    \includegraphics[width=0.1\textwidth]{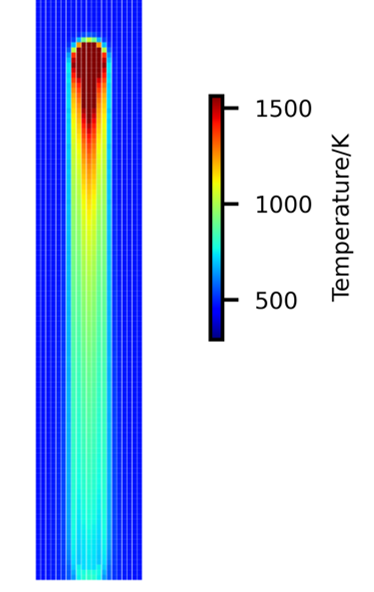}
    \includegraphics[width=0.45\textwidth]{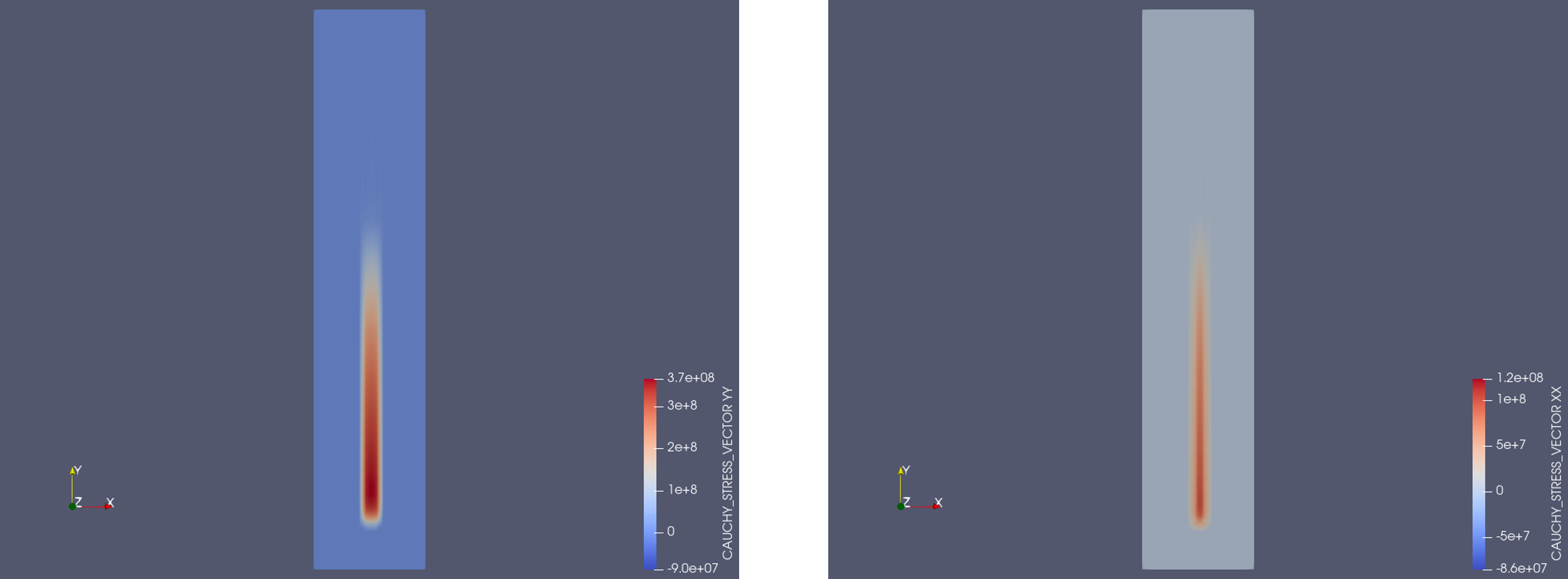}
    \caption{Snapshot of temperature distribution around the melt pool and the fine scale residual stress. The pixel size is 20 $\mu$m. The melt pool length is 0.32 mm. Left: stress along the scanning direction. Right: stress in the transverse direction.}
    \label{fig:fine_scale_result}
\end{figure}

The second step is to obtain the path-level thermal history. The thermal history, along with the ratio of the residual stress from the previous step, will be used to compute the effective thermal strain. We use PBF-CAPL to compute the path-level thermal history. The scanning paths are given in Figure \ref{fig:validation_path} and the process parameters are given in Figure \ref{table:validation_process_parameters}. 



The third step is to compute the effective thermal strain. In this step, we convert the results of the fine-scale problem into a model to be used in the path-level problem. Recall the purpose of the effective thermal strain is to reproduce the same anisotropic stresses. For the fine-scale problem, the residual stress comes from the anisotropic thermal gradient which causes non-uniform deformation around the melt pool. The deformation in the path-level simulation is more uniform since one path-level element covers the entire hatching space. To achieve the same anisotropic stresses, we alter the thermal strain on the path level from the conventional isotropic one to an effective anisotropic one (assuming scanning along $x$ direction): 
\begin{align}\label{eq:effective_thermal}
    \bm{\epsilon_{t,e}}(T) = (\int_{T_m}^T\alpha(t)dt) [1,r,1,0,0,0]^T
\end{align}
where $r$ leads to the anisotropic stresses. Now the question becomes how to determine $r$. Note deformation and plasticity happen mostly when temperature is high (stiffness is low) and stresses build up after cooling down. When stresses build up, we consider the deformation of the path-level element to be approximated to the plastic deformation, therefore $\epsilon_e + \epsilon_t = \epsilon_{total}-\epsilon_p = 0$, then by Hooke's law we have the following equation for a path-level element:
\begin{align}\label{eq:ratio}
    r = \frac{\epsilon_{t,x}}{\epsilon_{t,y}} = \frac{\epsilon_{e,x}}{\epsilon_{e,y}} = \frac{1/E(\sigma_x - \nu\sigma_y)}{1/E(\sigma_y-\nu\sigma_x)} = \frac{1-\nu\frac{\sigma_y}{\sigma_x}}{\frac{\sigma_y}{\sigma_x}-\nu}
\end{align}

Because we need to ensure the same anisotropic stresses, the ratio $\frac{\sigma_y}{\sigma_x}$ needs to be equal to the one from the fine-scale problem.  The corresponding value in the fine-scale problem is the ratio of the average stresses of the stress distributions around the melt pool $\frac{\Bar{\sigma}_y}{\Bar{\sigma}_x}$. Now $r$ is available when we substitute $\frac{\Bar{\sigma}_y}{\Bar{\sigma}_x} = \frac{\sigma_y}{\sigma_x}$ into equation \ref{eq:ratio}. Then we have the expression of the effective thermal strain in equation \ref{eq:effective_thermal}: its coefficient $\int_{T_m}^T\alpha(t)dt$ reflects the thermal history while the value $r$ causes the anisotropic stresses. The average residual stress is computed by taking all the elements' stress except for those whose magnitude is smaller than $\epsilon$. The exclusion of these elements is to exclude the non-melted elements. In the current paper we choose $\epsilon = 1$ KPa. The average residual stress is computed as $\Bar{\sigma_x} = 61$ MPa, $\Bar{\sigma_y} = 30$MPa, so we have the ratio $r = 0.2$ according to equation \ref{eq:ratio} and the effective thermal strain $\bm{\epsilon_{t,e}}(T) = (\int_{T_m}^T\alpha(t)dt) [1,0.2,1,0,0,0]^T$.

The last step is to apply the effective thermal strain in the path-level simulation. The path-level simulation use the same implementation as the conventional voxel-based approach except for two differences: (a) The thermal strain will be replaced by the effective thermal strain, and (b) the domain is discretized to capture the scanning path (instead of fine discretization to capture the melt pool).

Because we use the effective thermal strain and it is no longer needed to capture the details of the melt pool shape, the discretization is only needed to capture the scanning path instead of the melt pool shape. In our thermal simulation using PBF-CAPL, the discretization is linked to the scanning path, with the path-level elements having their widths set equal to the hatch space. In the mechanical simulation, the discretization is needed to capture the same scanning path. For simplicity, we use a regular voxel mesh whose element size is the hatch space since such a discretization is enough to capture the scanning path. The temperature of the mechanical element is mapped from the nearest thermal element. Compared to the conventional approaches, both our thermal and mechanical simulations have a much coarser discretization and thus are more computationally efficient. 

We compare the path-level simulation results with the results of voxel-based approach from literature~\cite{parry2016understanding}. The scanning of a single-layer Ti6Al4V powder is simulated. Scanning paths include a post-contour scanning of (a) unidirectional parallel scanning and (b) alternating parallel scanning. We conduct the path-level simulation of the same paths with our new proposed approach. As discussed previously, our approach only needs discretization resolution to capture the scanning paths, the element size is small as long as it is comparable to the PBF-CAPL element. Here, the element size is $90 \times 100 \times 40~mm^3$  (in the reference paper the element size is $20 \times 20 \times 20~mm^3$). It can be seen our approach uses much fewer elements ($20\times20\times20:90\times100\times40=1:45$).

\begin{figure}
    \centering
    \includegraphics[width=0.45\textwidth]{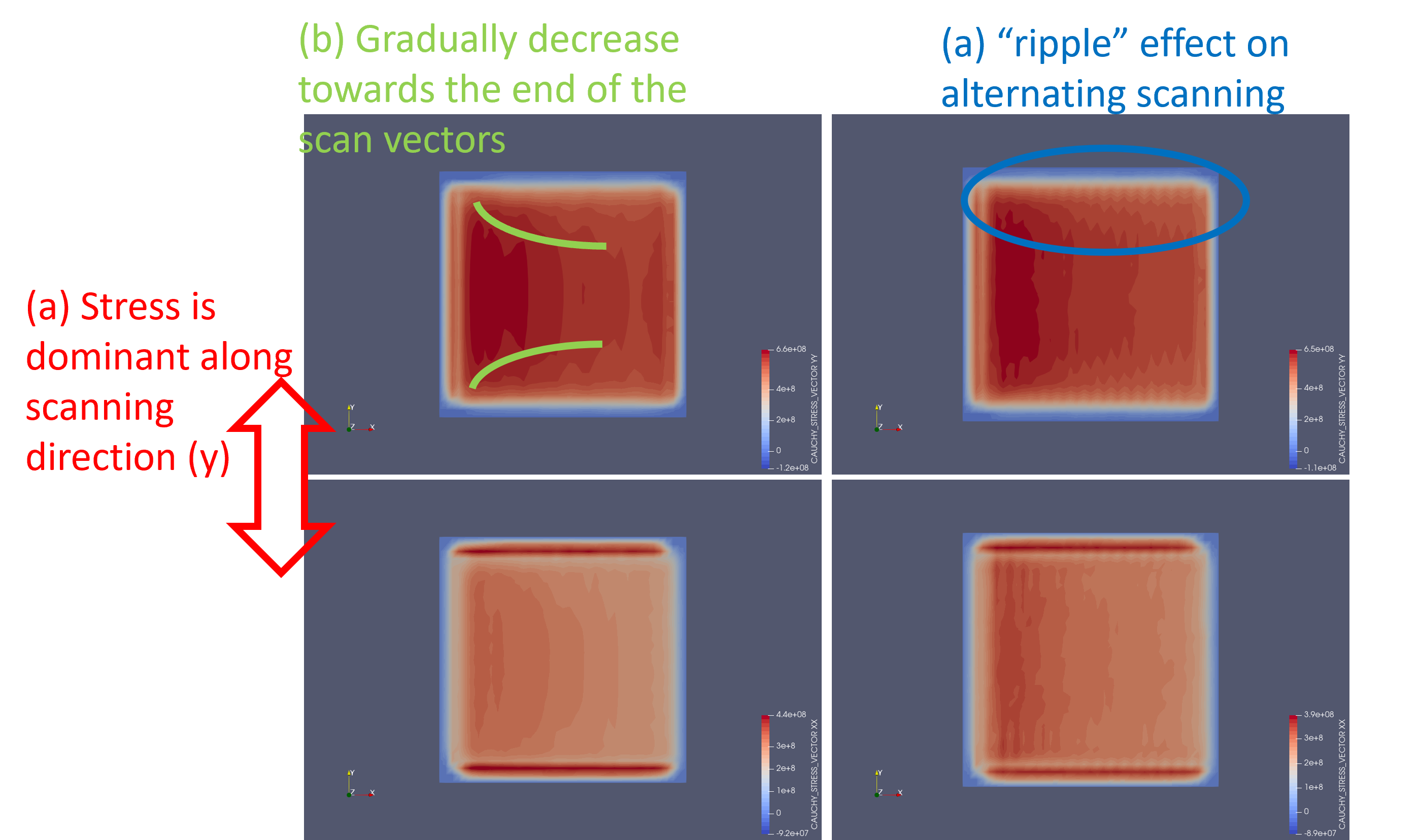}
    \caption{Simulation results by our approach (unidirectional at left and alternating at right). The top is the stress in y direction and the bottom is the stress in x direction.}
    \label{fig:validation_result}
\end{figure}

\ch{The results show that our approach is a good match compared to the voxel-based approach.} It is noted by the researcher that there are three scanning-path-led patterns~\cite{parry2016understanding}: (a) the "ripple effect" is found in the stress distribution, (b) stress is situated centrally along the hatch region and decreases toward the end of the scan vectors in the hatched region, and (c) the stress in the scanning direction is more dominant compared to those in the transverse direction. As shown in Figure \ref{fig:validation_result}, all these three patterns are replicated by our path-level simulation. The maximum magnitude of the stresses also approximately matches the results from the literature. The slight difference between these numbers might come from different physics models. \ch{For example, the preclusion of latent heat in the reference implementation lead to abnormally high maximum temperature. The inclusion of latent heat in PBF-CAPL significantly decrease in the maximal temperature into the normal range, result in a quantitatively different stress than the reference implementation. }

\begin{table}
\caption{Process parameters and material properties of Ti6Al4V~\cite{parry2016understanding}}\label{table:validation_process_parameters} 
\centering

\footnotesize
\begin{tabular}{ll}
 \hline
 Melting temperature &   1923 K \\
 Environmental temperature &   473 K \\
 Hatch space & 90 $\mu$m \\
 Layer thickness & 40 $\mu$m \\
 Laser power for hatching & 82.5 W \\
 Laser power for contour & 40 W \\
 Laser speed for hatching & 0.5 m/s\\
 Laser speed for contour & 0.25 m/s \\
 Environmental convection coefficient & 10 W/(m$^2$ K)\\
 Laser spot size & 50 $\mu$m
\end{tabular}
\end{table} 

\ch{We further demonstrate that our approach is indeed reasonable by comparing it to the key parameters used in the inherent strain method~\cite{liang2018modified}. The cross-validation results are available in another paper~\cite{liang2018modified}}, where the authors obtains the inherent strain for Ti6Al4V is determined as $\epsilon_x = 0.013$ and $\epsilon_y=0.003$. If we consider thermal history to be uniform with constant $r$, then our approach can be described as the inherent strain method on the path level. Considering the thermal expansion coefficient is around the order of $10^-5 /K$ and the melting point for the metal is around 1923 K, the thermal strain (cooling down from melting point to room temperature) can be estimated as $(1923-293)K*10^-5/K=0.0163$, which is approximated to the order of 0.013 given by the literature. Recalling we have the ratio $r=0.2$, the inherent strain is estimated to be $\epsilon_x = 0.0163$ and $\epsilon_y=0.0163*0.2 = 0.00326$, which is a good match with the results from the literature $\epsilon_y=0.003$. 

we further demonstrate that our approach is indeed reasonable by comparing it to the key parameters used in the inherent strain method~\cite{liang2018modified}. . Cross-validation results can be found in another paper~\cite{liang2018modified}, where the inherent strain for Ti6Al4V is determined as $\epsilon_x = 0.013$ and $\epsilon_y=0.003$. 

If we assume the thermal history to be uniform with a constant $r$, our approach can be conceptualized as the inherent strain method at the path level. Considering the thermal expansion coefficient is approximately $10^{-5}/K$ and the melting point of the metal is around 1923 K, the thermal strain (cooling from the melting point to room temperature) can be estimated as $(1923-293)K \times 10^{-5}/K = 0.0163$. This is approximately equivalent to the value of 0.013 given in the literature. Taking into account our ratio $r = 0.2$, the inherent strain is estimated to be $\epsilon_x = 0.0163$ and $\epsilon_y = 0.0163 \times 0.2 = 0.00326$. This estimation aligns well with the literature value of $\epsilon_y = 0.003$.

\section{Residual stresses on island patterns}

In this section, we will simulate the residual stresses of various island checkerboard patterns. We first conduct simulations of single islands as well as single layers that consist of multiple full and trimmed islands. We then discuss the influence of path-level thermal history and the layer boundaries on the residual stress with the simulation results. \ch{We observed two factors that affect the path-level residual stress: one is the the uneven shrinkage caused by the order of solidification, another is the thermal history more specifically the cooling rate. }

\subsection{Residual stresses on various single-island pattern}



\ch{Island patterns are increasingly popular for the LPBF scanning paths~\cite{jhabvala2010effect,chen2021island}.} In this paper, scanning patterns including unidirectional scanning, checkerboard patterns, and spiral patterns are simulated as single-layer tests, shown in Figure \ref{fig:table_of_islands}. The color from blue to red represents the laser start and end positions. We use the same process parameters for all the patterns, see table \ref{table:new_model_process_parameters}. All the patterns have the same 2 mm $\times$ 2 mm dimension.  
\begin{table}
\caption{Process parameters for island tests.}\label{table:new_model_process_parameters} 
\centering

\footnotesize
\begin{tabular}{ll}
 \hline
 Environmental temperature &   473 K \\
 Absorptivity & 0.77 \\
 Hatch space & 100 $\mu$ m \\
 Layer thickness & 40 $\mu$ m \\
 Laser power & 80 W \\
 Laser speed & 1 m/s \\
 Environmental convection coefficient & 10 W/(m$^2$ K)\\
 Laser spot diameter & 50 $\mu$m \\
 Platform thickness & 4 mm
\end{tabular}
\end{table}

\begin{table}[htbp]
  \centering
  \begin{minipage}[b]{0.15\linewidth}
    \centering
    \includegraphics[width=\linewidth]{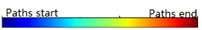}
    \label{fig:figure1}
  \end{minipage}
  \quad
  \begin{minipage}[b]{0.15\linewidth}
    \centering
    \includegraphics[width=\linewidth]{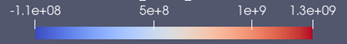}
    \label{fig:figure2}
  \end{minipage}
  \quad
  \begin{minipage}[b]{0.15\linewidth}
    \centering
    \includegraphics[width=\linewidth]{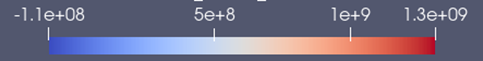}
    \label{fig:figure3}
  \end{minipage}
  \vspace{-0.3cm} 
  
  \begin{minipage}[b]{0.15\linewidth}
    \centering
    \includegraphics[width=\linewidth]{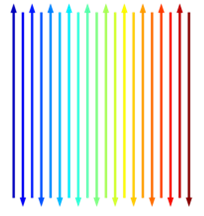}
    \label{fig:figure4}
  \end{minipage}
  \quad
  \begin{minipage}[b]{0.15\linewidth}
    \centering
    \includegraphics[width=\linewidth]{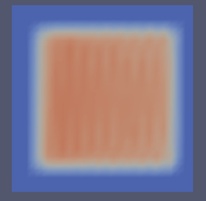}
    \label{fig:figure5}
  \end{minipage}
  \quad
  \begin{minipage}[b]{0.15\linewidth}
    \centering
    \includegraphics[width=\linewidth]{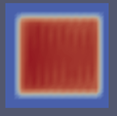}
    \label{fig:figure6}
  \end{minipage}
  \vspace{-0.3cm} 
  
  \begin{minipage}[b]{0.15\linewidth}
    \centering
    \includegraphics[width=\linewidth]{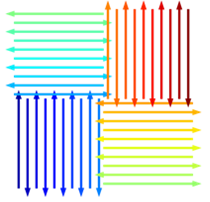}
    \label{fig:figure7}
  \end{minipage}
  \quad
  \begin{minipage}[b]{0.15\linewidth}
    \centering
    \includegraphics[width=\linewidth]{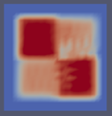}
    \label{fig:figure8}
  \end{minipage}
  \quad
  \begin{minipage}[b]{0.15\linewidth}
    \centering
    \includegraphics[width=\linewidth]{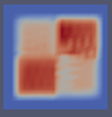}
    \label{fig:figure9}
  \end{minipage}
  \vspace{-0.3cm} 
  
  \begin{minipage}[b]{0.15\linewidth}
    \centering
    \includegraphics[width=\linewidth]{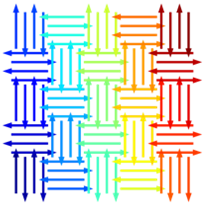}
    \label{fig:figure10}
  \end{minipage}
  \quad
  \begin{minipage}[b]{0.15\linewidth}
    \centering
    \includegraphics[width=\linewidth]{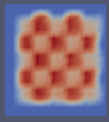}
    \label{fig:figure11}
  \end{minipage}
  \quad
  \begin{minipage}[b]{0.15\linewidth}
    \centering
    \includegraphics[width=\linewidth]{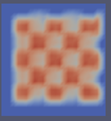}
    \label{fig:figure12}
  \end{minipage}
  \vspace{-0.3cm} 
  
  \begin{minipage}[b]{0.15\linewidth}
    \centering
    \includegraphics[width=\linewidth]{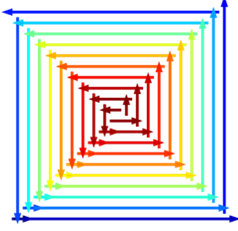}
    \label{fig:figure16}
  \end{minipage}
  \quad
  \begin{minipage}[b]{0.15\linewidth}
    \centering
    \includegraphics[width=\linewidth]{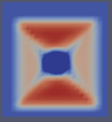}
    \label{fig:figure17}
  \end{minipage}
  \quad
  \begin{minipage}[b]{0.15\linewidth}
    \centering
    \includegraphics[width=\linewidth]{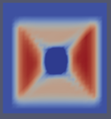}
    \label{fig:figure18}
  \end{minipage}
  \vspace{-0.3cm} 
  
  \begin{minipage}[b]{0.15\linewidth}
    \centering
    \includegraphics[width=\linewidth]{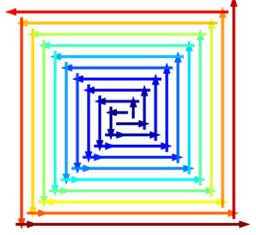}
    \label{fig:figure19}
  \end{minipage}
  \quad
  \begin{minipage}[b]{0.15\linewidth}
    \centering
    \includegraphics[width=\linewidth]{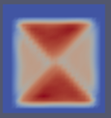}
    \label{fig:figure20}
  \end{minipage}
  \quad
  \begin{minipage}[b]{0.15\linewidth}
    \centering
    \includegraphics[width=\linewidth]{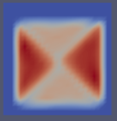}
    \label{fig:figure21}
  \end{minipage}
  \vspace{-0.3cm} 
  
  \begin{minipage}[b]{0.15\linewidth}
    \centering
    \includegraphics[width=\linewidth]{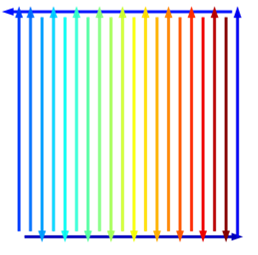}
    \label{fig:figure22}
  \end{minipage}
  \quad
  \begin{minipage}[b]{0.15\linewidth}
    \centering
    \includegraphics[width=\linewidth]{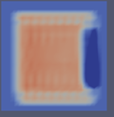}
    \label{fig:figure23}
  \end{minipage}
  \quad
  \begin{minipage}[b]{0.15\linewidth}
    \centering
    \includegraphics[width=\linewidth]{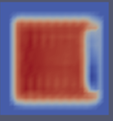}
    \label{fig:figure24}
  \end{minipage}
  \vspace{-0.3cm} 
  
  \begin{minipage}[b]{0.15\linewidth}
    \centering
    \includegraphics[width=\linewidth]{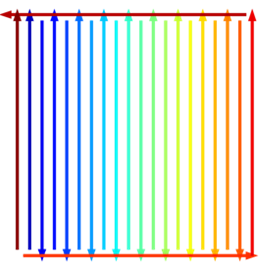}
    \label{fig:figure25}
  \end{minipage}
  \quad
  \begin{minipage}[b]{0.15\linewidth}
    \centering
    \includegraphics[width=\linewidth]{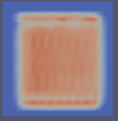}
    \label{fig:figure26}
  \end{minipage}
  \quad
  \begin{minipage}[b]{0.15\linewidth}
    \centering
    \includegraphics[width=\linewidth]{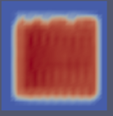}
    \label{fig:figure27}
  \end{minipage}
  
  \caption{Table of single island residual stress (Pa).}
  \label{fig:table_of_islands}
\end{table}

The residual stress shown in Figure \ref{fig:table_of_islands} are those when the layer is cooling down to environmental temperature. The simulation results show that the residual stress at a specific point is mostly dominated by the scanning direction through the specific point except for some regions in two cases: the inward spiral case and the pre-contour case. In most cases, the residual stress is tensile (positive) as the consequence of the volume is shrinking due to cooling: the materials tend to shrink, but the shrinking tendency is constrained by its neighbor. In other words, a specific point is stretched by its neighborhood, therefore the stress is tensile. However, there are compressive stress regions in the pre-contour and the inward spiral cases. In both two cases, the compressive regions are the last region to be scanned and solidified. The compressive stress is likely due to the already solidified neighborhoods which act as constraints that limit the deformation of the last region in all directions. Though the last region intends to shrink, it has expanded in the transverse direction due to the Poisson effect. Such expansive deformation is limited by the neighborhood which causes strong compression.

\subsection{Residual stresses on multiple trimmed islands}
\begin{figure}
    \centering
    \includegraphics[width=0.3\textwidth]{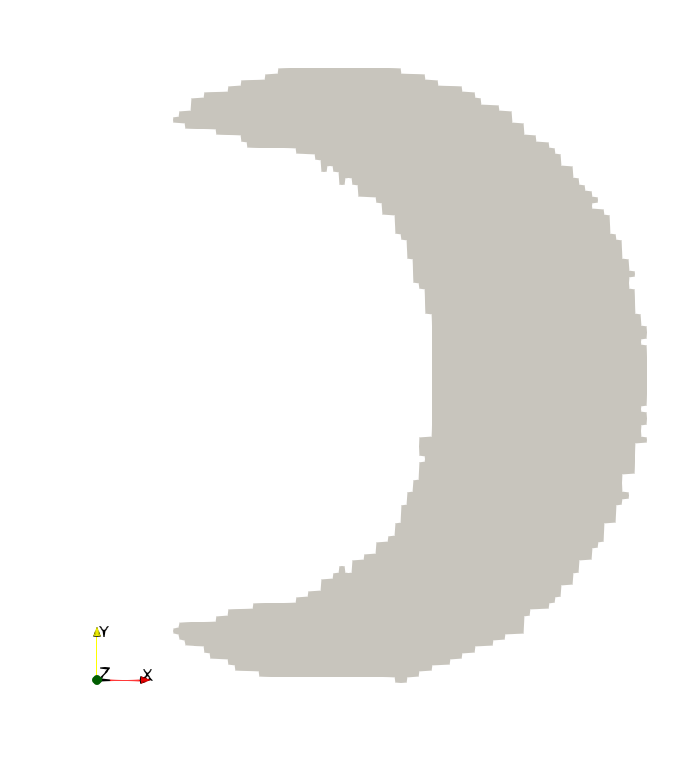}
    \caption{The crescent moon shape of the single layer simulation.}
    \label{fig:crescent-moon-shape}
\end{figure}

\ch{We simulate the LPBF process on a larger crescent moon shape which contains multiple full and trimmed islands.} As shown in Figure \ref{fig:crescent-moon-shape}, the scanning on a crescent moon of 8 mm $\times$ 10 mm is simulated. The process parameters are the same as those used in the single-island simulation. 

\begin{figure}
    \centering
    \includegraphics[width=0.9\textwidth]{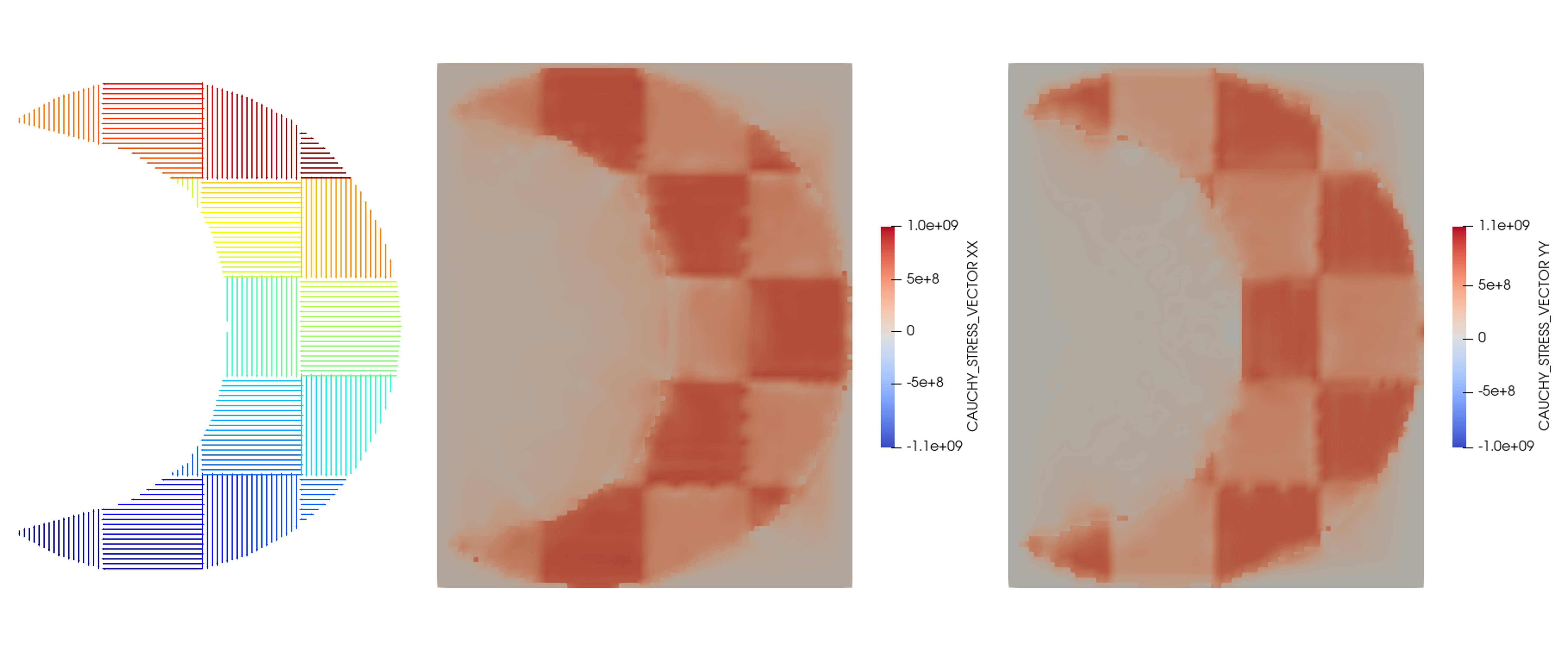}
    \caption{The Cauchy residual stress in XX and YY directions of an alternating parallel islands crescent moon.}
    \label{fig:alternating-90_stress}
\end{figure}

\begin{figure}  
    \centering
    \includegraphics[width=0.9\textwidth]{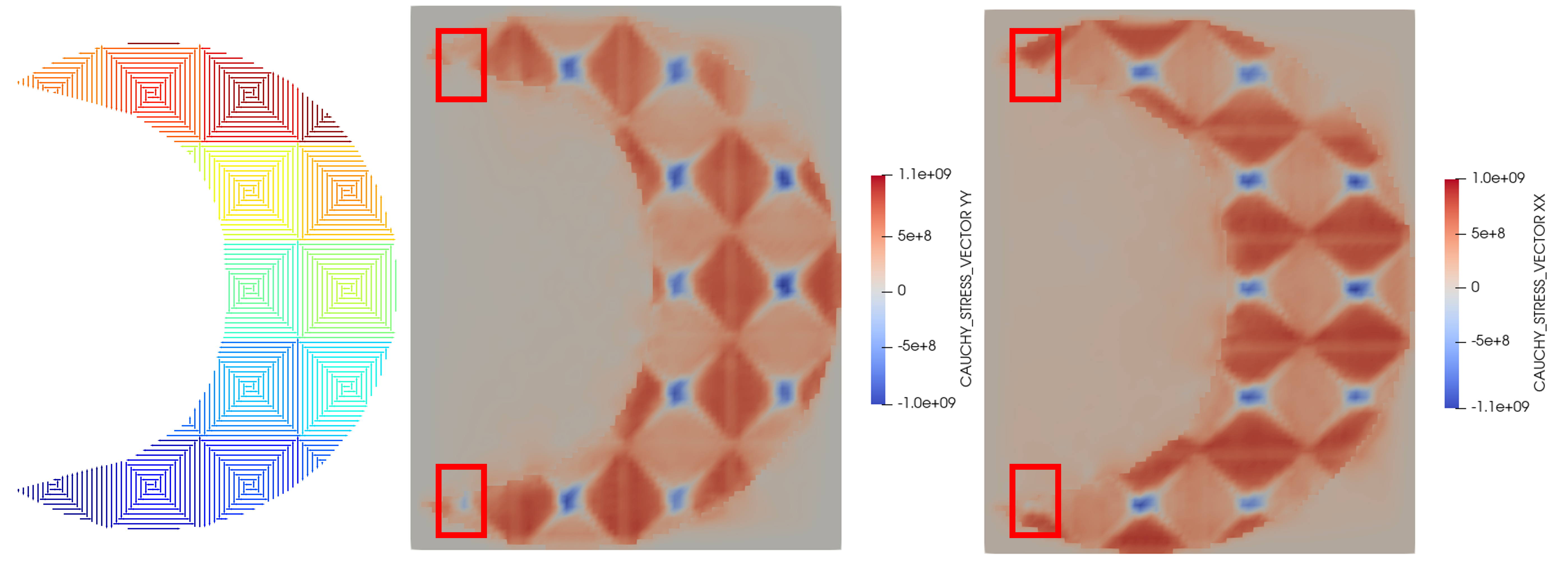}
    \caption{The Cauchy residual stress in XX and YY directions of an spiral islands crescent moon. The compressive region in the red boxes are not clearly visible due to they are on the layer shape boundary.}
    \label{fig:inf_stress}
\end{figure}

\ch{We tested our simulation approach on various island patterns. The simulation results indicate that the residual stress patterns generally resemble those from single island tests.} As shown in Figures \ref{fig:alternating-90_stress} and \ref{fig:inf_stress}, the results of most spiral islands are similar to the one in the single island test, except for the left bottom corner one in the red box. The compressive region is barely visible in the red box. \ch{We postulate the reason for this difference is that the center of the island is no longer constrained by its neighborhood in all directions since part of the island is trimmed off. Again, this case shows how the scanning order and the layer shape can affect the residual stress.}

\begin{figure}
    \centering
    \includegraphics[width=0.8\textwidth]{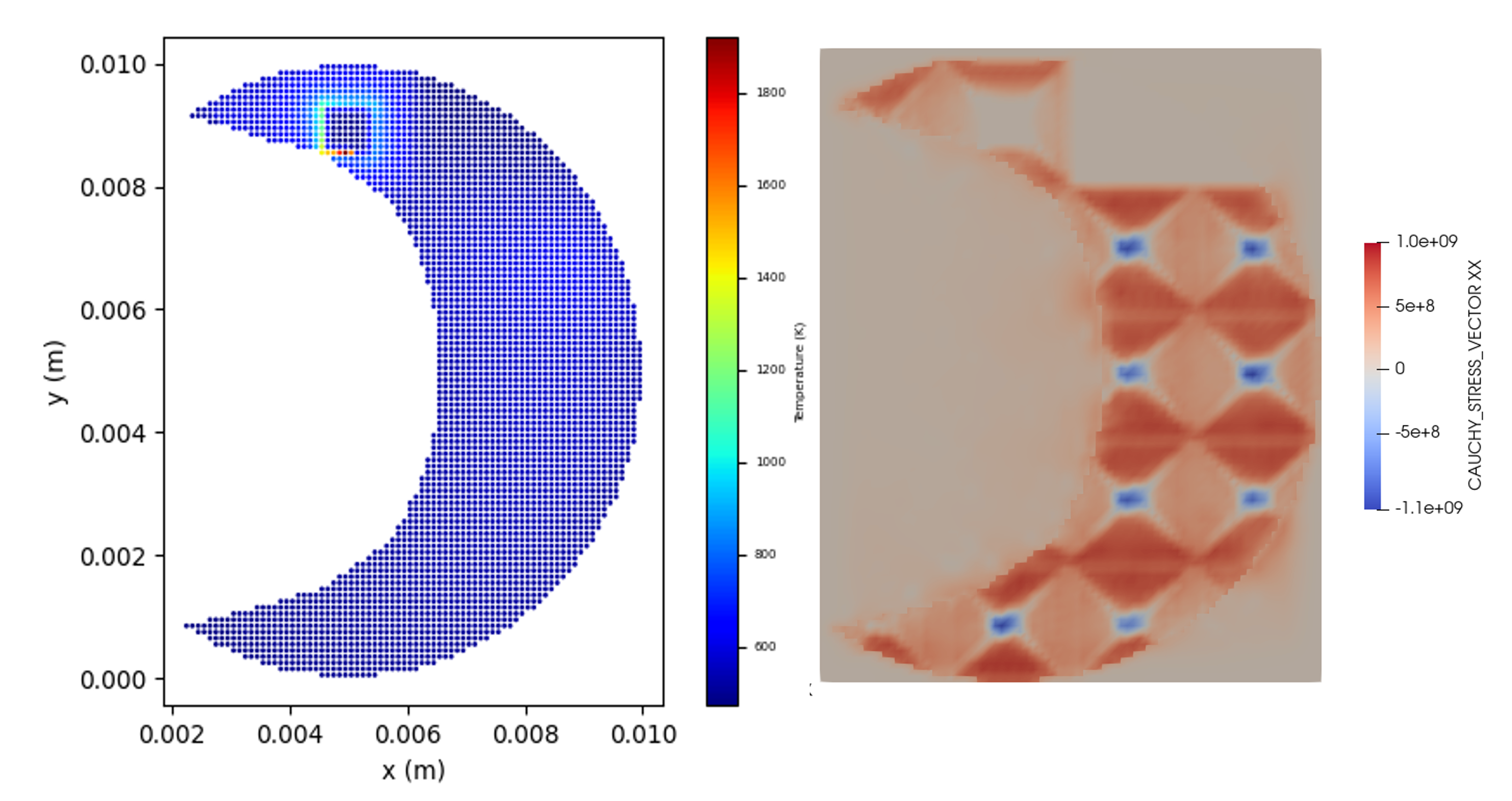}
    \caption{The residual stress and the temperature during the scanning process.}
    \label{fig:rapid_cooling}
\end{figure}

It is noteworthy that most of our islands have similar residual stress as those in the single-island tests. Note that $\epsilon_{total} = \epsilon_e + \epsilon_t+\epsilon_p$, such similarity suggests for those islands, their $\epsilon_{total}$ and $\epsilon_t$ are similar those on the other islands respectively. This similarity of $\epsilon_{total}$ is attributed to the fact that the deformation is small in the single-layer test. In the single-layer test, the layer is fully attached to the platform as there is no underlying overhang region, therefore deformation is expected to be small. \ch{ In general cases, small deformation is expected in the absence of overhangs or when overhangs are present but adequately supported.} The similarity of $\epsilon_t$ indicates the thermal history is similar on different islands because of the rapid cooling and solidification as the layer is fully connected to the underlying platform. Figure \ref{fig:rapid_cooling} demonstrates the stress in one island rapidly builds up as the laser moves away from this island: as the laser is scanning in the island at $x\in[0.004,0.006], y\in[0.008,0.01]$ mm, the temperatures are elevated and stress levels are low. Temperature rapidly decrease to below 1000 K and the stress increases in regions where the laser moves away from these areas.

\begin{figure}
    \centering
    \includegraphics[width=0.9\textwidth]{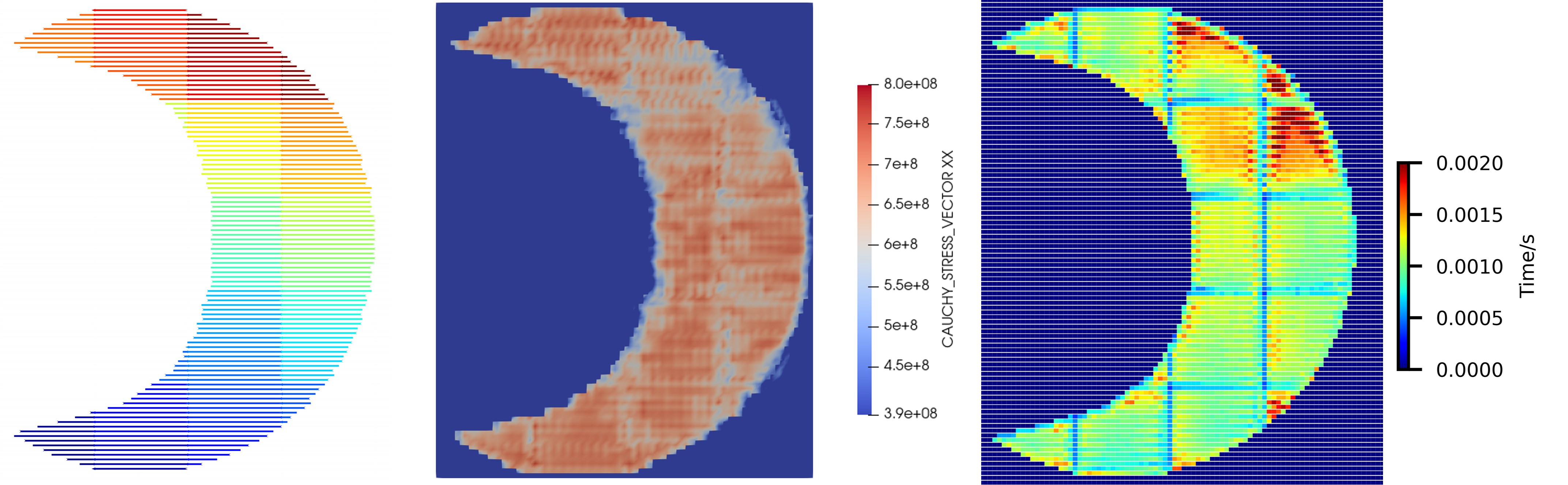}
    \caption{The residual stress and the time over threshold of parallel scanning paths.}
    \label{fig:parall_results}
\end{figure}

\begin{figure}
    \centering
    \includegraphics[width=0.9\textwidth]{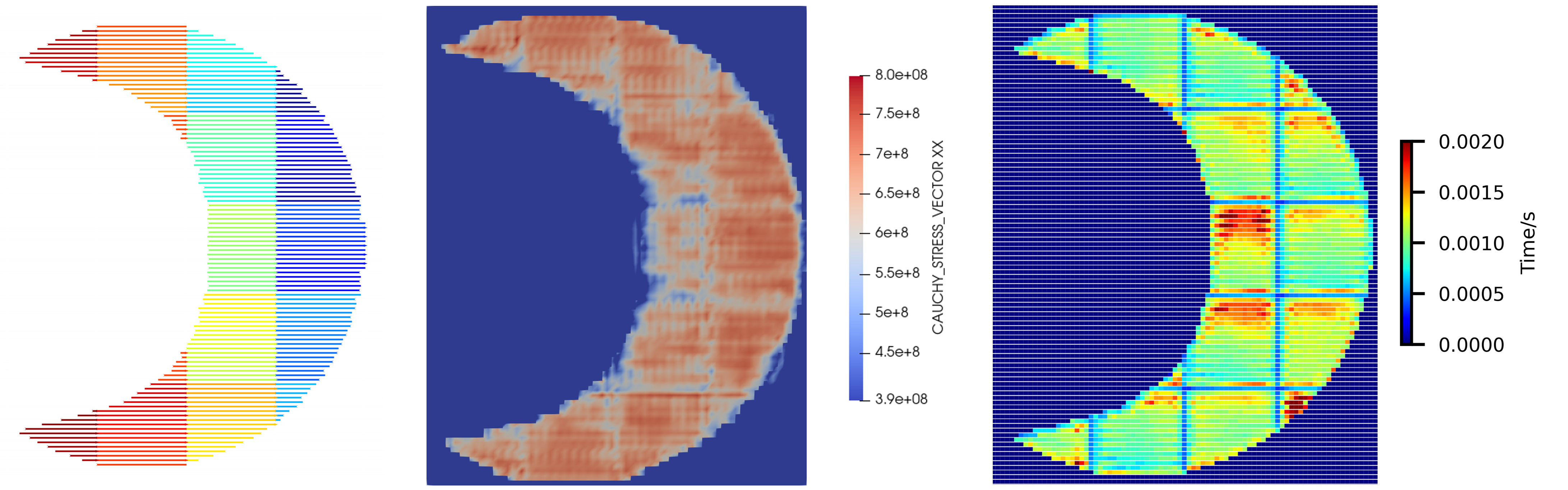}
    \caption{The residual stress and the time over threshold of parallel scanning paths in the reverse order of the case in Figure \ref{fig:parall_results}.}
    \label{fig:reverse_results}
\end{figure}

\begin{figure}
    \centering
    \includegraphics[width=0.9\textwidth]{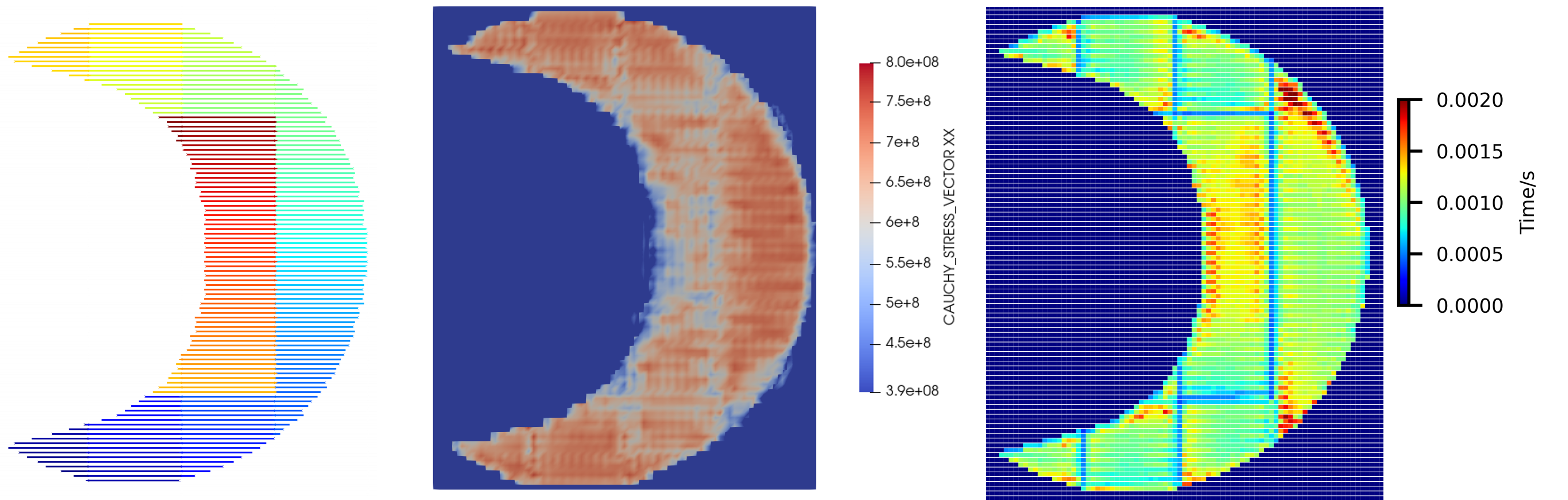}
    \caption{The residual stress and the time over threshold of parallel scanning paths in the spiral order of the case in Figure \ref{fig:parall_results}.}
    \label{fig:spiral_results}
\end{figure}

\ch{
Such highly localized stress formation results in similar stress distributions across different islands with the same pattern. Despite their similarity, the influence of heat accumulation on stress, resulting from the scanning paths, remains noticeable.} \ch{ We demonstrate such difference in Figures \ref{fig:parall_results} -\ref{fig:spiral_results}. We use the time over threshold temperature~\cite{lane2022statistical} to demonstrate the difference of thermal histories.} Here we choose the threshold to be 923 K. \ch{It shows that the thermal history varies among different islands based on their order, owing to differences in thermal accumulation.} \ch{ The time over threshold is larger for the islands that are scanned after their neighboring islands, compared to those scanned earlier in the sequence. In other words, the cooling for these islands are slower. } \ch{Compared with the stress distribution, it is shown that the low residual stress regions coincide with the region having a higher time over threshold. This indicate the slow slow cooling is likely to cause less residual stress.} These cases show how the sequence of islands can affect the cooling rate which causes different residual stress distribution.

\section{Conclusion and future directions}
\label{sec:conclusion}

In this paper, we propose an approach to simulate the residual stress on the path level. \ch{The proposed approach is capable to utilize the path-level thermal history to simulate the residual stress. } The proposed approach is capable of capturing the path-level thermal and mechanical evolution. Compared with existing approaches, the present approach is more accurate than the layer-based approach since it is capable to simulate on the path level, and it is more efficient than the conventional voxel-based approaches as the new approach does not need to capture the details of the melt pool. \ch{We validated our present approach by comparing our simulation results with the results by the voxel-based approach from the literature. A good match is obtained.} We discussed the factors that affect the residual stress by conducting numerical simulations with our approach to the island's patterns. The influence of the path-level thermal history and the layer shape on the residual stress are discussed. 

Understanding the gaps in existing methods is crucial for advancing the simulation of LPBF processes. Currently, available approaches provide efficient layer-wise methods like the inherent strain method, along with accurate but more expensive voxel-based approaches. While the efficient layerwise methods excel in regions where the path is less significant, identifying the regions where the path plays a crucial role and determining how to address them remain challenges. It is imperative to comprehend the assumptions or process parameters that dictate the efficacy of these methods.

The stress observations from our study shed light on these challenges. We found that the stress in the single-island test closely resembles that of the multi-island layer. This similarity might be attributed to the localized formation of residual stress, a result of rapid cooling and minimal deformation in the single-layer test. We demonstrated that our approach can be simplified to a path-level inherent strain method with these observation assumed. However, we also highlighted that stress can be influenced by the complex layer boundary and the sequence of solidification.

Incorporating path-level stress simulation and the path-level thermal simulation could be a pivotal component of a comprehensive part-scale LPBF simulation strategy. We have previously developed a hybrid layer-path thermal simulation approach \cite{}. The path-level simulation is conducted at the region where the scanning path has critical influence. The mechanical simulation methodology introduced in this paper could potentially serve as a foundation for a comparable layer-path mechanical simulation, bridging the identified gaps in existing methods.
\begin{appendices}
\setcounter{secnumdepth}{0}
\section{Appendix: PBF-CAPL}

In our previous work, we developed a path-level thermal simulation framework for the LPBF process called PBF-CAPL~\cite{liu2024scalable}. The PBF-CAPL is developed based on a Contact-aware path-level (CAPL) discretization approach was developed for simulating the path-level thermal history of a moving heat source ~\cite{zhang2018linear,zhang2019towards,zhang2022scalable}. CAPL tailors discretization to the manufacturing toolpath and adopts locality for linear time complexity in part-scale thermal history simulations. We validate the new approach against melt pool images captured with the co-axial melt pool monitoring (MPM) system on the Manufacturing Metrology Testbed (AMMT) developed at the National Institute of Standards and Technology (NIST)~\cite{lane2016design, yeung2020residual}. Excellent agreements in the length and width of melt pools are found between simulations and experiments conducted on a custom-controlled laser powder bed fusion (LPBF) testbed on a nickel-alloy (IN625) solid surface. A brief introduction is given below for PBF-CAPL. The readers might find more details in our previous paper~\cite{liu2024scalable}.

There are two main stages in PBF-CAPL: the pre-processing stage and the execution stage. In the pre-processing stage, there is a path-level discretization algorithm that discretizes the simulation domain into scanning-path-associated elements. The elements are connected by a data structure called a contact graph. In the execution stage, a lumped-capacitance heat transfer model is used to simulate the moving-laser-driven convection-conduction-radiation problem. The linear complexity is achieved by utilizing a concept we call "active body": a region that is “close” to the current heat source either in time or space. Only elements in the active body are updated in each time step. The lazy update is applied for the region outside the active body.

In the preprocessing stage, the data structure is prepared for the execution stage. In the path-level discretization algorithm, the scan path and laser power information are extracted from the input file. Along with these real paths, fictitious paths are added to fill the entire domain if it is needed. The real and fictitious paths are discretized into elements. PBF-CAPL considers an element to be associated with a sub-path of the real or fictitious path. Each element is approximated as a path-aligned box defined by its length $L$ (equal to the length of the sub-path), width $W$, and height $H$ (equal to the layer height). The cross-sectional shape is a rectangle $W\times H$ which is perpendicular to the scanning direction. The top surface of the element is approximated by a rectangle $L\times H$. The width of an element $W$ is determined by a Voronoi diagram of the real and fictitious paths. For parallel scanning lines, $W$ will be simply the hatching space. The elements are connected by the contact graph. The contact graph is composed of vertices which represent elements and edges which represent the contacts between adjacent elements. A contact will be generated if two elements overlap.

In the execution stage, a lumped-capacitance heat transfer model is used to simulate the thermal process. The equation for an element $i$ is:
\begin{align}
    m_ic_i\Dot{T}_i = Q_i + q_i
\end{align}
where $m_i$, $c_i$, $T_i$ are the mass, thermal capacity, and the temperature of element $i$, respectively. $Q_i$ and $q_i$ are the heat transfer on the boundary of element $i$ and the laser input energy on element $i$, respectively. $Q_i$ further includes the convection from the environment, radiation to the environment, and the conduction from other elements.

In each step, computation based on the lumped-capacitance model is conducted in the "active body". The active body is a region near the laser. The active body moves along with the laser. The size of the active body is determined based on trial and experience. A proper size of the active body ensures the cooling outside the active body the analytical Newton's cooling law is applicable. Utilizing the active body ensures the computation is only needed in the active body in each step. The thermal history outside the active body needs to be updated by Newton's cooling only when the elements go into the active body again.  
\end{appendices}

\section*{Acknowledgements}
This research was supported by the National Institute of Standards and Technology. The responsibility for errors and omissions lies solely with the author. 



\section*{References}

\bibliography{ref}

\begin{thebibliography}{10}
\expandafter\ifx\csname url\endcsname\relax
  \def\url#1{\texttt{#1}}\fi
\expandafter\ifx\csname urlprefix\endcsname\relax\def\urlprefix{URL }\fi
\expandafter\ifx\csname href\endcsname\relax
  \def\href#1#2{#2} \def\path#1{#1}\fi

\bibitem{vock2019powders}
S.~Vock, B.~Kl{\"o}den, A.~Kirchner, T.~Wei{\ss}g{\"a}rber, B.~Kieback, Powders for powder bed fusion: a review, Progress in Additive Manufacturing 4~(4) (2019) 383--397.

\bibitem{bhavar2017review}
V.~Bhavar, P.~Kattire, V.~Patil, S.~Khot, K.~Gujar, R.~Singh, A review on powder bed fusion technology of metal additive manufacturing, Additive manufacturing handbook (2017) 251--253.

\bibitem{singh2021powder}
D.~D. Singh, T.~Mahender, A.~R. Reddy, Powder bed fusion process: A brief review, Materials Today: Proceedings 46 (2021) 350--355.

\bibitem{liu2016homogenization}
X.~Liu, V.~Shapiro, Homogenization of material properties in additively manufactured structures, Computer-Aided Design 78 (2016) 71--82.

\bibitem{luthi2023adaptive}
C.~L{\"u}thi, M.~Afrasiabi, M.~Bambach, An adaptive smoothed particle hydrodynamics (sph) scheme for efficient melt pool simulations in additive manufacturing, Computers \& Mathematics with Applications 139 (2023) 7--27.

\bibitem{moges2019review}
T.~Moges, G.~Ameta, P.~Witherell, A review of model inaccuracy and parameter uncertainty in laser powder bed fusion models and simulations, Journal of manufacturing science and engineering 141~(4) (2019) 040801.

\bibitem{patil2021benchmark}
N.~Patil, R.~Ganeriwala, J.~M. Solberg, N.~E. Hodge, R.~M. Ferencz, Benchmark multi-layer simulations for residual stresses and deformation in small additively manufactured metal parts, Additive Manufacturing 45 (2021) 102015.

\bibitem{liang2018modified}
X.~Liang, L.~Cheng, Q.~Chen, Q.~Yang, A.~C. To, A modified method for estimating inherent strains from detailed process simulation for fast residual distortion prediction of single-walled structures fabricated by directed energy deposition, Additive Manufacturing 23 (2018) 471--486.

\bibitem{liang2021incorporating}
X.~Liang, W.~Dong, Q.~Chen, A.~C. To, On incorporating scanning strategy effects into the modified inherent strain modeling framework for laser powder bed fusion, Additive Manufacturing 37 (2021) 101648.

\bibitem{liu2024scalable}
X.~Liu, X.~Liu, N.~G. Kumar, P.~Witherell, Scalable path level thermal history simulation of powder bed fusion process validated by melt pool images, Additive Manufacturing (2024) 104111.

\bibitem{parry2016understanding}
L.~Parry, I.~Ashcroft, R.~D. Wildman, Understanding the effect of laser scan strategy on residual stress in selective laser melting through thermo-mechanical simulation, Additive Manufacturing 12 (2016) 1--15.

\bibitem{denlinger2017thermomechanical}
E.~R. Denlinger, M.~Gouge, J.~Irwin, P.~Michaleris, Thermomechanical model development and in situ experimental validation of the laser powder-bed fusion process, Additive Manufacturing 16 (2017) 73--80.

\bibitem{chen2019effect}
C.~Chen, J.~Yin, H.~Zhu, Z.~Xiao, L.~Zhang, X.~Zeng, Effect of overlap rate and pattern on residual stress in selective laser melting, International Journal of Machine Tools and Manufacture 145 (2019) 103433.

\bibitem{zhang2020scanning}
W.~Zhang, M.~Tong, N.~M. Harrison, Scanning strategies effect on temperature, residual stress and deformation by multi-laser beam powder bed fusion manufacturing, Additive Manufacturing 36 (2020) 101507.

\bibitem{gourdin1986dynamic}
W.~H. Gourdin, Dynamic consolidation of metal powders, Progress in Materials Science 30~(1) (1986) 39--80.

\bibitem{queva2020numerical}
A.~Queva, G.~Guillemot, C.~Moriconi, C.~Metton, M.~Bellet, Numerical study of the impact of vaporisation on melt pool dynamics in laser powder bed fusion-application to in718 and ti--6al--4v, Additive Manufacturing 35 (2020) 101249.

\bibitem{bidare2018fluid}
P.~Bidare, I.~Bitharas, R.~Ward, M.~Attallah, A.~J. Moore, Fluid and particle dynamics in laser powder bed fusion, Acta Materialia 142 (2018) 107--120.

\bibitem{matthews2017denudation}
M.~J. Matthews, G.~Guss, S.~A. Khairallah, A.~M. Rubenchik, P.~J. Depond, W.~E. King, Denudation of metal powder layers in laser powder-bed fusion processes, in: Additive Manufacturing Handbook, CRC Press, 2017, pp. 677--692.

\bibitem{michopoulos2018multiphysics}
J.~G. Michopoulos, A.~P. Iliopoulos, J.~C. Steuben, A.~J. Birnbaum, S.~G. Lambrakos, On the multiphysics modeling challenges for metal additive manufacturing processes, Additive Manufacturing 22 (2018) 784--799.

\bibitem{ganeriwala2021towards}
R.~K. Ganeriwala, N.~E. Hodge, J.~M. Solberg, Towards improved speed and accuracy of laser powder bed fusion simulations via multiscale spatial representations, Computational Materials Science 187 (2021) 110112.

\bibitem{luo2018survey}
Z.~Luo, Y.~Zhao, A survey of finite element analysis of temperature and thermal stress fields in powder bed fusion additive manufacturing, Additive Manufacturing 21 (2018) 318--332.

\bibitem{cao2021novel}
Y.~Cao, X.~Lin, N.~Kang, L.~Ma, L.~Wei, M.~Zheng, J.~Yu, D.~Peng, W.~Huang, A novel high-efficient finite element analysis method of powder bed fusion additive manufacturing, Additive Manufacturing 46 (2021) 102187.

\bibitem{yang2019residual}
Y.~Yang, M.~Allen, T.~London, V.~Oancea, Residual strain predictions for a powder bed fusion inconel 625 single cantilever part, Integrating Materials and Manufacturing Innovation 8~(3) (2019) 294--304.

\bibitem{denlinger2015residual}
E.~R. Denlinger, J.~C. Heigel, P.~Michaleris, Residual stress and distortion modeling of electron beam direct manufacturing ti-6al-4v, Proceedings of the Institution of Mechanical Engineers, Part B: Journal of Engineering Manufacture 229~(10) (2015) 1803--1813.

\bibitem{sow2020influence}
M.~Sow, T.~De~Terris, O.~Castelnau, Z.~Hamouche, F.~Coste, R.~Fabbro, P.~Peyre, Influence of beam diameter on laser powder bed fusion (l-pbf) process, Additive Manufacturing 36 (2020) 101532.

\bibitem{grunewald2021influence}
J.~Gr{\"u}newald, F.~Gehringer, M.~Schm{\"o}ller, K.~Wudy, Influence of ring-shaped beam profiles on process stability and productivity in laser-based powder bed fusion of aisi 316l, Metals 11~(12) (2021) 1989.

\bibitem{yan2015multiscale}
W.~Yan, J.~Smith, W.~Ge, F.~Lin, W.~K. Liu, Multiscale modeling of electron beam and substrate interaction: a new heat source model, Computational Mechanics 56~(2) (2015) 265--276.

\bibitem{dunbar2016experimental}
A.~J. Dunbar, E.~R. Denlinger, M.~F. Gouge, P.~Michaleris, Experimental validation of finite element modeling for laser powder bed fusion deformation, Additive Manufacturing 12 (2016) 108--120.

\bibitem{liu2022understanding}
J.~Liu, G.~Li, Q.~Sun, H.~Li, J.~Sun, X.~Wang, Understanding the effect of scanning strategies on the microstructure and crystallographic texture of ti-6al-4v alloy manufactured by laser powder bed fusion, Journal of Materials Processing Technology 299 (2022) 117366.

\bibitem{nadammal2021critical}
N.~Nadammal, T.~Mishurova, T.~Fritsch, I.~Serrano-Munoz, A.~Kromm, C.~Haberland, P.~D. Portella, G.~Bruno, Critical role of scan strategies on the development of microstructure, texture, and residual stresses during laser powder bed fusion additive manufacturing, Additive Manufacturing 38 (2021) 101792.

\bibitem{chen2021prediction}
C.~Chen, Z.~Xiao, Y.~Wang, X.~Yang, H.~Zhu, Prediction study on in-situ reduction of thermal stress using combined laser beams in laser powder bed fusion, Additive Manufacturing 47 (2021) 102221.

\bibitem{peng2018fast}
H.~Peng, M.~Ghasri-Khouzani, S.~Gong, R.~Attardo, P.~Ostiguy, R.~B. Rogge, B.~A. Gatrell, J.~Budzinski, C.~Tomonto, J.~Neidig, et~al., Fast prediction of thermal distortion in metal powder bed fusion additive manufacturing: Part 2, a quasi-static thermo-mechanical model, Additive Manufacturing 22 (2018) 869--882.

\bibitem{bayat2020part}
M.~Bayat, C.~G. Klingaa, S.~Mohanty, D.~De~Baere, J.~Thorborg, N.~S. Tiedje, J.~H. Hattel, Part-scale thermo-mechanical modelling of distortions in laser powder bed fusion--analysis of the sequential flash heating method with experimental validation, Additive Manufacturing 36 (2020) 101508.

\bibitem{ueda1979new}
Y.~Ueda, K.~Fukuda, M.~Tanigawa, New measuring method of three dimensional residual stresses based on theory of inherent strain (welding mechanics, strength \& design), Transactions of JWRI 8~(2) (1979) 249--256.

\bibitem{dadvand2010object}
P.~Dadvand, R.~Rossi, E.~O{\~n}ate, An object-oriented environment for developing finite element codes for multi-disciplinary applications, Arch Computat Methods Eng 17 (2010) 253--297.
\newblock \href {http://dx.doi.org/10.1007/s11831-010-9045-2} {\path{doi:10.1007/s11831-010-9045-2}}.

\bibitem{dadvand2013migration}
P.~Dadvand, R.~Rossi, M.~Gil, X.~Martorell, J.~Cotela, E.~Juanpere, S.~Idelsohn, E.~O{\~n}ate, Migration of a generic multi-physics framework to hpc environments, Computers \& Fluids 80 (2013) 301--309.
\newblock \href {http://dx.doi.org/10.1016/j.compfluid.2012.02.004} {\path{doi:10.1016/j.compfluid.2012.02.004}}.

\bibitem{ferrandiz2022kratos}
V.~M. Ferrándiz, P.~Bucher, R.~Zorrilla, R.~Rossi, J.~Cotela, A.~C. Velázquez, M.~A. Celigueta, J.~Maria, T.~Teschemacher, C.~Roig, M.~Maso, G.~Casas, S.~Warnakulasuriya, M.~Núñez, P.~Dadvand, S.~Latorre, I.~de~Pouplana, J.~I. González, F.~Arrufat, J.~Gárate, {KratosMultiphysics/Kratos: Release 9.2 (v9.2)}, \url{https://doi.org/10.5281/zenodo.3234644} (2022).
\newblock \href {http://dx.doi.org/10.5281/zenodo.3234644} {\path{doi:10.5281/zenodo.3234644}}.

\bibitem{jhabvala2010effect}
J.~Jhabvala, E.~Boillat, T.~Antignac, R.~Glardon, On the effect of scanning strategies in the selective laser melting process, Virtual and physical prototyping 5~(2) (2010) 99--109.

\bibitem{chen2021island}
Q.~Chen, H.~Taylor, A.~Takezawa, X.~Liang, X.~Jimenez, R.~Wicker, A.~C. To, Island scanning pattern optimization for residual deformation mitigation in laser powder bed fusion via sequential inherent strain method and sensitivity analysis, Additive Manufacturing 46 (2021) 102116.

\bibitem{lane2022statistical}
B.~Lane, H.~Yeung, Z.~Yang, Statistical and spatio-temporal data features in melt pool monitoring of additive manufacturing, in: IIE Annual Conference. Proceedings, Institute of Industrial and Systems Engineers (IISE), 2022, pp. 1--6.

\bibitem{zhang2018linear}
Y.~Zhang, V.~Shapiro, Linear-time thermal simulation of as-manufactured fused deposition modeling components, Journal of Manufacturing Science and Engineering 140~(7).

\bibitem{zhang2019towards}
Y.~Zhang, V.~Shapiro, P.~Witherell, Towards thermal simulation of powder bed fusion on path level, in: International Design Engineering Technical Conferences and Computers and Information in Engineering Conference, Vol. 59179, American Society of Mechanical Engineers, 2019, p. V001T02A034.

\bibitem{zhang2022scalable}
Y.~Zhang, V.~Shapiro, P.~Witherell, A scalable framework for process-aware thermal simulation of additive manufacturing processes, Journal of Computing and Information Science in Engineering 22~(1).

\bibitem{lane2016design}
B.~Lane, S.~Mekhontsev, S.~Grantham, M.~Vlasea, J.~Whiting, H.~Yeung, J.~Fox, C.~Zarobila, J.~Neira, M.~McGlauflin, et~al., Design, developments, and results from the nist additive manufacturing metrology testbed (ammt), in: 2016 International Solid Freeform Fabrication Symposium, University of Texas at Austin, 2016.

\bibitem{yeung2020residual}
H.~Yeung, B.~Lane, A residual heat compensation based scan strategy for powder bed fusion additive manufacturing, Manufacturing letters 25 (2020) 56--59.

\end{thebibliography}

\end{document}